# A multi-stage looking-ahead holding strategy to stabilize a high-frequency bus line

Sheng-Xue He[1][*]

1. Business School, University of Shanghai for Science and Technology, Shanghai 200093, China;

**Abstract**: If a bus line becomes unstable, passengers' waiting time will be lengthened and buses' capacities will be mismatched. To stabilize a high-frequency bus line, many holding strategies have been proposed. Among these strategies, some need to take oversimplified assumptions to simplify the formulation of a real bus line; some may choose a myopic holding time because they take into account only the local and currently available headway information. To overcome the above shortcomings, we proposed an adaptive holding strategy which continuously reduces the deviations of the instantaneous headways from the Dynamic Target Headway (DTH). By using DTH, the new strategy can make use of the global headway information to determine a proper holding time. To fully estimate the influence of a holding time on the operation of a bus line in a relative long time period, we introduced the multi-stage looking-ahead mechanism into our new strategy. A detailed bus line simulation system was used to realize the above mechanism and to describe all kinds of complex components of a bus line. The numerical experiment demonstrated the effectiveness of our new strategy. Some meaningful insights were uncovered as follow: a). The number of stages to be looked ahead should not be too small or too big; b). The bigger the range of the action set, the better the performance of our strategy; c). To obtain the optimal performance, the interval between two neighbor holding times should not be too small or too big; d). the bigger the number of the control points, the better the performance of our strategy.

## 1. Introduction

The unevenly dispersed buses along a bus route commonly appear in a high-frequency bus line. Sometimes the unevenly dispersed buses will evolve into bus bunching. The negative influences of the above phenomenon on passengers mainly include the lengthened waiting times and the congested riding experiences. The decreased service level and the worsened reliability of bus line service will further reduce public transit ridership. To avoid the above problems, we will propose an adptive holding strategy to stabilize a high-frequency bus line in this paper.

Researchers have proposed many effective ways to improve the operational stability and reliability of a bus line over the past several decades. The representative ways include the stop-skipping strategies (Suh et al. 2002; Fu et al. 2003; Sun and Hickman 2005; Cortés et al. 2010; Liu et al. 2013), limited-boarding strategies (Osuna and Newell 1972; Newell 1974; Barnett 1974; Delgado et al., 2009, 2012), embedding-slack strategies (Daganzo 1997a, b; Zhao et al. 2006; Daganzo 2009a; Xuan et al. 2011), static and dynamic holding strategies (Hickman 2001; Eberlein et al. 2001, Sun and Hickman 2008; Puong and Wilson 2008; Daganzo 2009a; Xuan et al. 2011; Bartholdi and Eisenstein 2012; Delgado et al. 2012; He 2015; Argote-Cabanero et al. 2015; He et al. 2019a; Liang et al. 2019 a, b), speed adjustment strategies (Daganzo 2009b; Daganzo and Pilachowski 2009, 2011; He et al. 2019b), transit signal priority strategies (Liu et al. 2003; Ling and Shalaby 2004; Estrada et al. 2016), bus substitution strategy (Petit et al. 2018), the real-time holding to reconcile transfer synchronization and service regularity (Gavriilidou and Cats, 2018) and the real-time bus dispatching strategies ( Berrebi et al. 2015; Cristobal et al. 2018; Huang et al. 2019; Luo et al. 2019).

The headway-based holding strategies have been studied by many researchers. Based on the basic control theory, Daganzo (2009a) formulated a control model to determine holding times for buses at control points. The control objective in Daganzo (2009a) is to minimize the differences between the actual arrivals of buses and a given schedule or a virtual schedule during the whole observation period. Later many holding strategies (including Xuan et al. 2011; He 2015; Argote-Cabanero et al. 2015; Petit et al. 2018) were proposed based on the fundamental model of Daganzo (2009a). Two weak points lie in the above strategies. On the one hand, since a high-frequency bus line usually has no pre-specified schedule or headway, the above holding strategies cannot be applied to an above bus line. On the other hand, to formulate their optimal control models, these strategies usually need some oversimplified assumptions to describe the complex bus line. For example, signalized intersections are seldom considered in their formulations.

To stabilize a bus line without any pre-specified schedule or headway, researchers proposed some rule-based holding strategies (Bartholdi and Eisenstein 2012; Liang et al. 2016, 2019a, b; Zhang and Lo 2018; He et al. 2019a). Bartholdi and Eisenstein (2012) presented a holding strategy which only uses the backward headway information at the terminal station and does not require any scheduled or virtual headway as reference. Liang et al. (2016 and 2019a, b) proposed to equalize the forward and backward headways at terminal control points. Zhang and Lo (2018) presented a theoretical framework to investigate the properties of two-way headway-based holding strategies. He et al. (2019a) proposed a holding strategy to stabilize a bus line with dynamic target headway. These strategies have been proven very effective in some cases in practice. But because these strategies used only the local and short-term headway information to choose a holding time, the long-term influence of the chosen holding time on the operation stability of a bus line may be negative.

By simulating a real bus line in detail, a strategy can avoid taking some oversimplified assumptions to describe the bus line. In other words, a strategy based on a detailed simulation system can effectively cope with all kinds of real situations, e.g. the movements of vehicles at a signalized intersection. To avoid making a myopic decision, a holding strategy should not only make use of the global information of a bus line but also take into account the long-term influence of the chosen holding time on the operational stability of a bus line.

In view of the above reflection, we will try to design an adaptive holding strategy which will avoid the above problems. Firstly, we will define the Dynamic Target Headway (DTH) as the average of instantaneous headways associated with a given time point. Based on DTH, we will further define not only a stability index to measure the operational stability of a bus line, but also the action cost to measure the influence of a chosen holding time on the operational stability. The global headway information is captured in DTH. To take into account the long-term influence of a control action in the decision making, we will introduce the multi-stage looking-ahead mechanism into our strategy. To realize the above mechanism, we will adopt a detailed simulation system as the basis to estimate various expected values. By doing so, the operation of a real bus line can be described in detail. A detailed test bus line will be constructed to demonstrate the effectiveness of our strategy. We also expect to uncover some meaningful insights into the application of our strategy.

The remainder of this paper is organized as follows. Section 2 introduces control objective and control action. Section 3 presents the multi-stage looking-ahead holding strategy. Section 4 demonstrates the effectiveness of our strategy with a detailed test bus line. Section 5 summarizes the main contents and points out several future research directions.

## 2. Control Objective and Control Action

### 2.1. Control Objective

*2.1.1. Dynamic Target Headway*

For a high-frequency bus line without any pre-specified schedule or headway, we need to find some criterion to decide on the specific holding time. In this subsection, we will define Dynamic Target Headway (DTH). The chosen holding time should reduce the difference between DTH and the headway of the current bus.

We will employ a circular bus line to demonstrate our method. The DTH associated with time $t$ is defined as follows:

$$H(t) = \sum_{b \in B} h_b(t) / n_B, \qquad (1)$$

where $h_b(t)$ is the time headway of bus $b$ to its nearest leading bus at time $t$, $B$ is the set of buses, and $n_B$ is the total number of buses. $h_b(t)$ stands for the travel time required for bus $b$ from its current position to the current position of its nearest leading bus. DTH will be used later as a target to adjust the headway of the current bus by holding it at its current stop for a specified time interval.

If the operation state of a bus line remains steady during the observation period, we can figure out a fixed ideal headway as the target to adjust the headway of the current bus. The Expected System Headway (ESH) is widely used in this situation. ESH can be calculated as follows:

$$\tilde{H} = (X / \tilde{v} + \sum_{i \in I} t_i^W + \sum_{e \in E} t_e^D) / n_B, \qquad (2)$$

where $X$ is the length of the bus line; $\tilde{v}$ is the average bus cruising speed; $t_i^W$ is the expected delay for a bus at intersection $i \in I$ where $I$ is the set of all the intersections; $t_e^D$ is the expected dwell time for a bus at stop $e$. $\tilde{H}$ is a fixed value for a bus line and so can be calculated once for all.

*2.1.2. Measure Stability*

With DTH associated with time $t$, we can defined its pseudo standard deviation as follows:

$$\sigma_H(t) = \sqrt{\sum_{b \in B} (h_b(t) - H(t))^2 / n_B}, \qquad (3)$$

This deviation can be used to evaluate the operational stability of a bus line at time instant $t$. A small value of $\sigma_H(t)$ means that all the instantaneous headways $h_b(t)$, $\forall b \in B$ are similar to each other. A big value of $\sigma_H(t)$ means headways are unevenly distributed along the bus line.

To survey a dynamically evolving system, we can choose some Critical Time Points (CTPs) to note down its performance instead of surveying it all the time continuously. In this study, we will check the whole operation situation of a bus line when a bus is about to depart from a stop after the necessary loading and discharging operations. After the above checking, a proper holding time can be decided for the bus. In view of the above operation, we will use the time when a bus about to depart from a stop after the necessary boarding and alighting as a CTP.

The process that the bus line evolves from one CTP to the next CTP is called 'rolling forward one stage'. The time interval between two successive CTPs is called a stage. A stage will be indicated by its starting time, i.e. the preceding CTP.

For simplicity, a typical stage is denoted by its starting CTP $t_m$. All the CTPs in a given observation time period $T$ constitute a set of stages denoted by $\bar{T}$. Suppose that there are total $n_{\bar{T}}$ stages in an observation period $T$. Due to the influence of various stochastic factors on the system, the value of $n_{\bar{T}}$ is unknown until the end of the observation period.

$\sigma_H(t_m)$ can be used to measure the operational stability at CTP $t_m$. To measure the whole operational stability of a bus line during an observation period, we define the following stability index of a bus line:

$$\bar{c}_H = (\sum_{t_m \in \bar{T}} \sigma_H(t_m))/n_{\bar{T}}. \tag{4}$$

The corresponding standard deviation of $\bar{c}_H$ is

$$\sigma_{\bar{c}} = \sqrt{[\sum_{t_m \in \bar{T}} (\sigma_H(t_m) - \bar{c}_H)^2]/(n_{\bar{T}} - 1)}. \tag{5}$$

From the above definitions, we can see that $\bar{c}_H$ and $\sigma_{\bar{c}}$ have the same measurement unit as time. $\bar{c}_H$ stands for the average deviation of headways from DCH over the whole observation period. $\sigma_{\bar{c}}$, i.e. the standard deviation of $\bar{c}_H$, shows the reliability of $\bar{c}_H$ as an expected estimate. $\bar{c}_H$ is the stability index to be used to evaluate the bus line's operation. Since $\bar{c}_H$ is defined as the average value over the cruising buses and the set $\bar{T}$, it intrinsically possesses the potential to indicate the operational stability of any bus line system as a whole.

The goal of our holding strategy is to minimize $\bar{c}_H$ against all the uncertain and stochastic factors influencing the operation of a bus line. It is easy to see that the smaller $\bar{c}_H$, the more steady and reliable the operation of the bus line during the observation period.

## 2.2. Control Action

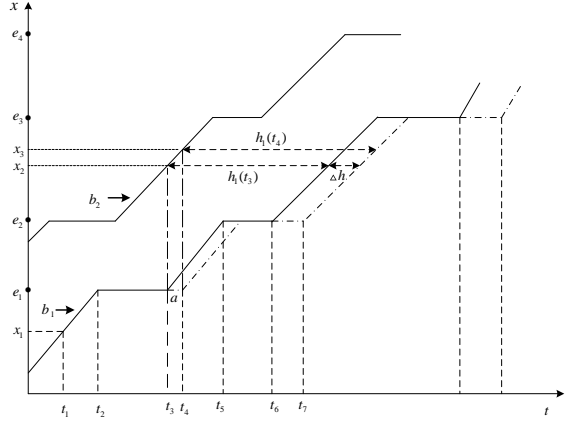

Fig. 1 Trajectories of two buses.

Figure 1 shows partial trajectories of two buses successively cruising in a bus line. Suppose that the current time is $t_3$. At this time, bus $b_1$ just finishes its loading and discharging operations at stop $e_1$. There are two alternative trajectories one of which bus $b_1$ can choose to follow at this time. Bus $b_1$ may directly depart from its current stop and follows the expected trajectory indicated by the solid slant line. Or we can hold this bus at its current stop for a time period $a = t_4 - t_3$ and then release it. Due to the holding operation, bus $b_1$ will follow the new trajectory indicated by the dash-dot slant line.

To hold a bus at its stop for a specified time period is the only control means considered in this study. The holding time $a$ is also called control action for convenience. To make the holding operation more practical, we assume that the value of $a$ can only be chosen from a given discrete set $A \equiv \{0, a_1, a_2, \ldots, a_{n_A}\}$ where $a_n = n\tau$ for $n \in \{0, 1, 2, \ldots, n_A\}$ and $\tau > 0$ is a given short time interval, such as 2 or 5 seconds. Obviously, we can specify a control action set for a given stop. In other words, the control action sets for two different stops may be different.

In real life, generally only part of stops will be used as control points. We will call a stop used as a control point a controllable stop. If a stop is not used as control point, we call it a non-control stop. For a non-control stop, we can assume that its only control action is $a = 0$ and its action set is the set $\{0\}$. By assigning the control action set $\{0\}$ to all the non-control stops, we can deal with all the stops including controllable and non-control stops in the same way when deploying our strategy.

Holding a bus for a while at a stop will change the time headway of the bus to its leading bus.

As shown in Figure 1, due to the holding operation $a$, the time headway between bus $b_1$ and its leading bus $b_2$ changes from $h_1(t_3)$ to $h_1(t_4)$. By choosing a proper holding time with our strategy, we hope that the resulted headway change will be helpful for stabilizing the bus line in the long run.

## 3. Multi-Stage Looking-ahead Holding Strategy

The strategy to be presented in Section 3.2 will use the global headway information collected from multiple stages to select a proper holding time. The underlying idea of our strategy is to roll the simulation system forward several stages to see which control action, that is the holding time, will give rise to the highest operational stability of the simulated bus line. To roll the system forward, we need to note down the main features of the bus line system that is the state variable to be introduced in the following subsection.

### 3.1. State Variable

Suppose we are considering a stage indicated by a CTP $t_m$. A state variable is a series of features of a bus line associated with a given time point. The state variable of a bus line should have the potential to reproduce the trajectories of buses with the help of the actual control actions. So we will choose three features of a bus line to construct the state variable.

At CTP $t_m$, a bus may stop as a bus stop or is heading to a stop. This stop where the bus stops or is heading to is called the **target stop** associated with the bus at $t_m$. We choose target stops as the first feature used to constitute a state variable. $e_b^m$ is used to denote the target stop of bus $b$ with respect to the stage $t_m$. The second feature is the **latest arrival time** with respect to bus stops. $t_e^{\mathbb{A},m}$ is used to denote the latest arrive time of buses at stop $e$ with respect to the stage $t_m$. The third feature is the time interval associated with a bus between the current time and the departure time of the associated bus from its target stop. This will be called **the time interval for a bus to be activated.** $t_b^{\mathbb{D},m}$ is used to denote the time interval for bus $b$ to be activated after the stage $t_m$. Here 'to be activated' means to consider whether a holding operation needs to be applied on the bus in question after the given time interval.

Group all the above three features for all buses and stops at CTP $t_m$ into a vector denoted by $s_m$. Vector $s_m$ is the state variable with respect to stage $t_m$. Use Figure 1 as an example to demonstrate the above conceptions. Assume that the current time is $t_1$. The current stage $t_m$ with respect to $b_1$ is $t_3$. $e_{b_1}^m$ and $t_{b_1}^{\mathbb{D},m}$ are $e_1$ and $t_3 - t_1$, respectively. If assume that the current time is $t_3$, $t_{e_1}^{\mathbb{A},m}$ will be $t_2$.

The constructed state variable can be used to reproduce the trajectories of buses with the help of the actual control actions. To reproduce the trajectory for a bus, we need to know its arrival and departure times from the bus stops in a proper order. According to the related definition, the arrival times of the bus at stops have been noted down by the second feature included in the state variable. If we add up the current time, the actual holding time, and the third feature included in the current state variable regarding to the current bus, we can obtain the departure time of the current bus from the current stop. With the obtained arrival and departure times, the trajectories of buses can be reproduced easily.

### 3.2. Multi-Stage Look Ahead Procedure

*3.2.1. Cost of Action*

When a holding time is determined, its impact on the stability of the bus line system needs to be measured so as to judge the quality of the decision. As mentioned earlier, we can use $\sigma_H(t)$ defined in Eq. (3) to assess the instantaneous operational stability of a bus line. Similar to $\sigma_H(t)$, we define the cost of action $a_i \in A_e$ with respect to stage $t_m$ as follows

$$c(a_i) = \sum_{b \in B}(h_b(a_i, t_m) - H(t_m))^2, \qquad (6)$$

where $h_b(a_i, t_m)$ is the forward time headway of bus $b$ at CTP $t_m$ when the holding time $a_i$ is chosen for the current bus. $H(t_m)$ stands for the DTH at stage $t_m$ before any holding time is chosen for the current bus. If we replace $h_b(t_m)$ in Eq. (3) with $h_b(a_i, t_m)$, we can see that $c(a_i)$ has the same changing trend as the modified $\sigma_H(t)$. So minimizing $c(a_i)$ is equal to minimizing the modified $\sigma_H(t_m)$.

When a non-control point is considered, the only action at such a stop is to dispatch the bus at once after the necessary boarding and alighting processes. In this situation, Eq. (6) will still be used to calculate the influence of some actions earlier determined. Based on the above observation, we should view $c(a_i)$ as a general conception to assess the operational stability of a bus line at CTP $t_m$ when action $a_i$ (sometimes $a_i$ stands for no action) is chosen.

*3.2.2. The Procedure of Multi-stage Look-ahead Strategy*

To avoid making a myopic decision and to make use of the global headway information embedded in the DTH at the same time, we will choose a holding time by looking several stages ahead from the current stage. The impact of the chosen holding time on the operational stability of the bus line can be assessed more accurately through looking several stages ahead. To look several stages ahead is to roll forward several stages of the simulation system of the bus line. To roll the simulation system forward, we will replace the actual or sample values of many random variables with the corresponding expected values. The random variables mainly include the travel time in a road segment, the delay at an intersection, and the dwell time of a bus at a stop. The actual or sample values of the above random variables will be used to simulate the actual running of a bus line during a given observation period.

We clarify two conceptions to be used later here. In this study, we call the part of bus route between two successive stops a bus line segment. We view a stop or an intersection as a critical point of the bus line in question. The part of bus route between two successive critical points is called a road segment. Obviously, a bus line segment may include several road segments and several intersections.

Assume that the current time is $\vec{t}$, the current system state is $s_m$ and the number of successive stages to be looked ahead is $N$. Some new notations are required in the following look-ahead procedure. $e_b^i$ stands for the bus stop for bus $b$ to be activated when the current looking ahead stage is $i \in \{1, 2, \ldots N\}$. $t_b^{D,i}$ is the remaining time for bus $b$ to be activated at the looking ahead stage $i \in \{1, 2, \ldots N\}$. $t_e^{la,i}$ is the modified latest arrival time associated with bus stop $e$ when the current looking ahead stage is $i \in \{1, 2, \ldots N\}$. Note that we use the time point of $\vec{t}$ as the origin to define the modified latest arrival time. To distinguish the expected value from the actual or sample value, an over bar will be used to indicate the expected value. So $\bar{t}_g^b$ stands for the expected travel time of bus $b$ in bus line segment $g$. $\bar{t}_e^b$ is the expected dwell time of bus $b$ at stop $e$.

The procedure of searching the optimal holding time $a^*$ by looking $N$ stage ahead is given as follows:

**Step 1**: Assign a large positive value to $\tilde{c}_i, \forall i \in \{1, 2, \cdots, N\}$.

**Step 2**: Choose a bus $b^1 \in \arg\min_{d \in B}\{t_d^{\mathbb{D},m}\}$ to be active at the first level. If the target stop $e_{b^1}^m$ is a controllable stop, go to Step 3; or else, let 0 be the optimal action $a^*$ and go to Step 4.

**Step 3:** Carry out the following multiple nested "for" loops.

**For** every action $a^1 \in A_{e_{b^1}^m}$, execute the following operations (start the first level):

$e_{b^1}^1 := e_{b^1}^m \oplus 1$ and $e_b^1 := e_b^m, \forall b \neq b^1$;

$t_{b^1}^{D,1} := t_{b^1}^{\mathbb{D},m} + a^1 + \bar{t}_{g^1}^{b^1} + \bar{t}_{e_{b^1}^1}^{b^1}$ and $t_b^{D,1} := t_b^{\mathbb{D},m}, \forall b \neq b^1$;

$$t^{l\,a^1}_{e^1_{b^1}} := t^{\mathbb{D},m}_{b^1} + a^1 + \overline{t}^1_{g^1} \text{ and } t^{la,1}_e := t^{\mathbb{A},m}_e - \vec{t}, \quad \forall e \neq e^1_{b^1}.$$

Note that the two ends of bus line segment $g^1$ are $e^m_{b^1}$ and $e^1_{b^1}$.

Calculate the cost of action $c(a^1)$ by eq. (6).

Choose a bus $b^2 \in \arg\min_{d \in B}\{t^{D,1}_d\}$ to be activated at the second level.

**For** every $a^2 \in A_{e^1_{b^2}}$, execute the following operations (start the second level):

$$e^2_{b^2} := e^1_{b^2} \oplus 1 \text{ and } e^2_b := e^1_b, \forall b \neq b^2;$$

$$t^{D,2}_{b^2} := t^{D,1}_{b^2} + a^2 + \overline{t}_{g^2} + \overline{t}^{b^2}_{e^2_{b^2}} \text{ and } t^{D,2}_b := t^{D,1}_b, \forall b \neq b^2;$$

$$t^{la,2}_{e^2_{b^2}} := t^{D,1}_{b^2} + a^2 + \overline{t}_{g^2} \text{ and } t^{la,2}_e := t^{la,1}_e, \quad \forall e \neq e^2_{b^2}.$$

Calculate the cost of action $c(a^2)$ by eq. (6).

…

Choose a bus $b^N \in \arg\min_{d \in B}\{t^{D,N-1}_d\}$ to be activated at the level $N$.

**For** every $a^N \in A_{e^{N-1}_{b^N}}$, execute the following operations (start the level $N$):

$$e^N_{b^N} := e^{N-1}_{b^N} \oplus 1 \text{ and } e^N_b := e^{N-1}_b, \forall b \neq b^N;$$

$$t^{D,N}_{b^N} := t^{D,N-1}_{b^N} + a^N + \overline{t}_{g^N} + \overline{t}^{b^N}_{e^N_{b^N}} \text{ and } t^{D,N}_b := t^{D,N-1}_b, \forall b \neq b^N;$$

$$t^{la,N}_{e^N_{b^N}} := t^{D,N-1}_{b^N} + a^N + \overline{t}_{g^N} \text{ and } t^{la,N}_e := t^{la,N-1}_e, \forall e \neq e^N_{b^N}.$$

Calculate the cost of action $c(a^N)$ by eq. (6);

Let $\tilde{c}_N := \min\{\tilde{c}_N, c(a^N)\}$.

**End** the "for" of the level $N$.

Let $\tilde{c}_{N-1} := \min\{\tilde{c}_{N-1}, c(a^{N-1}) + \gamma\tilde{c}_N\}$.

…

Let $\tilde{c}_2 := \min\{\tilde{c}_2, c(a^2) + \gamma\tilde{c}_3\}$.

**End** the "for" of the level 2.

Let $\tilde{c}_1 := \min\{\tilde{c}_1, c(a^1) + \gamma\tilde{c}_2\}$ and $a^* := \arg\min_{a^1}\{\tilde{c}_1, c(a^1) + \gamma\tilde{c}_2\}$

**End** the "for" of the level 1.

**Step 4**: Output the optimal action $a^*$ with respect to the current state $s_m$.

In the above process, $\gamma \in (0,1]$ is a given parameter used to discount the future expected cost. Note that in the above multi-stage looking-ahead procedure, the corresponding action set may be {0} at some stages. In this situation, the corresponding "for" loops will degenerate into simply carrying out the rolling operation without any holding operation.

## 4. Numerical Experiment

### 4.1. The Test Bus Line

In this section, we will present the detailed information about a test bus line so that interested readers can rebuild this bus line in detail.

*4.1.1. Buses*

We will use a circular bus line with 9 buses and 30 bus stops to demonstrate our approach. The length of the bus line is 17.95km. There are 13 intersections in the bus route.

Table 1

The basic data about the buses.

| No. of Bus | 1 | 2 | 3 | 4 | 5 | 6 | 7 | 8 | 9 |
|---|---|---|---|---|---|---|---|---|---|
| Passenger Capacity | 72 | 70 | 80 | 60 | 72 | 60 | 72 | 80 | 60 |
| Initial Target Stop | 1 | 4 | 8 | 11 | 15 | 18 | 21 | 25 | 28 |
| RTBA (s) | 20 | 0 | 40 | 30 | 50 | 10 | 30 | 35 | 25 |

Table 1 presents the main information about buses. The passenger capacity, the initial target stops where buses are dwelling at or are heading to at the beginning of simulation, and the Remaining Time for a bus to Be Activated at the beginning of the observation period (RTBA) are presented. From the data in the row indicated by "Initial Target Stop", we can see that at the beginning of the observation period the buses are evenly dispersed along the bus line.

*4.1.2. Bus Line Segment*

Table 2

The relations between bus line segments and road segments.

| BLS | RS | Length | BLS | RS | Length | BLS | RS | Length |
|---|---|---|---|---|---|---|---|---|
| 1 | 1,2 | 200,400 | 11 | 15 | 600 | 21 | 30,31 | 200,400 |
| 2 | 3 | 500 | 12 | 16,17 | 300,400 | 22 | 32 | 500 |
| 3 | 4 | 600 | 13 | 18,19 | 300,320 | 23 | 33 | 600 |
| 4 | 5,6 | 260,350 | 14 | 20 | 500 | 24 | 34,35 | 260,350 |
| 5 | 7 | 530 | 15 | 21 | 450 | 25 | 36 | 530 |
| 6 | 8 | 560 | 16 | 22,23,24 | 200,250,100 | 26 | 37 | 560 |
| 7 | 9 | 600 | 17 | 25 | 570 | 27 | 38 | 600 |
| 8 | 10,11 | 300,500 | 18 | 26 | 610 | 28 | 39,40 | 300,500 |
| 9 | 12 | 600 | 19 | 27 | 600 | 29 | 41 | 600 |
| 10 | 13,14 | 300,350 | 20 | 28,29 | 300,350 | 30 | 42,43 | 300,350 |

Note. 'BLS' and 'RS' stand for bus line segment and road segment, respectively.

Table 2 shows the inclusion relations between the road segments and the bus line segments. Data in the columns titled with BLS and RS are the serial numbers of the bus line segments and the corresponding road segments, respectively. The lengths of road segments are given in the column titled with 'Length'.

We assume that the average cruising speed of buses is 36(km/hr). But to mimic the actual running of a bus line, we need to consider the uncertainty of travel times. Assume that the length of road segment $d$ is $l_d$ (measured by meters). With the average cruising speed $\bar{v}$ in this road segment, the expected travel time $\bar{t}_d$ (measured by seconds) in $d$ is $l_d/\bar{v}$. Let $\delta_d$ be a normal random variable with mean 0 and variance $\sigma_d^2$. A sample travel time in $d$ can be generated from a random variable $t_d = \bar{t}_d + \delta_d$. The obtained sample travel time will be used as the actual travel time in our study. To simplify the computation, we assume that $\sigma_d = 0.005 l_d$ (measured by seconds) holds.

Assume that the signal control schemes at all the intersections are the pre-timed two phases with respect to the approach of the bus line in question. Table 3 lists the traffic signal cycle $t_i^{cl}$, the red phase $t_i^{red}$, the green phase $t_i^{green}$, and the remaining time of the initial phase $t_i^{or}$ for all intersections. Note that $t_i^{cl} = t_i^{red} + t_i^{green}$, $\forall i \in I$ holds. The row labeled by "Initial phase" specifies the initial phases of all the intersections. In this row, '1' and '2' indicate the red and green phases, respectively. The last row points out the bus line segment which includes the corresponding intersection.

Table 3

The basic data of intersections.

| No. of Intersection | 1 | 2 | 3 | 4 | 5 | 6 | 7 | 8 | 9 | 10 | 11 | 12 | 13 |
|---|---|---|---|---|---|---|---|---|---|---|---|---|---|
| $t_i^{red}$ (s) | 40 | 40 | 40 | 30 | 30 | 40 | 40 | 30 | 30 | 40 | 40 | 40 | 30 |
| $t_i^{green}$ (s) | 50 | 30 | 35 | 45 | 30 | 30 | 45 | 35 | 45 | 50 | 30 | 35 | 45 |
| $t_i^{or}$ (s) | 20 | 20 | 20 | 20 | 20 | 20 | 30 | 20 | 20 | 10 | 20 | 10 | 20 |
| Initial phase | 2 | 1 | 1 | 2 | 2 | 1 | 2 | 2 | 2 | 2 | 1 | 1 | 2 |
| Bus line segment | 1 | 4 | 8 | 10 | 12 | 13 | 16 | 16 | 20 | 21 | 24 | 28 | 30 |

The expected delay at intersection $i$ is given by $\frac{1}{2}(t_i^{red})^2/t_i^{cl}$. This expected delay will be used in the multi-stage looking-ahead strategy. The actual delay at an intersection will be mimicked as follows. When a bus arrives at an intersection and the encountered traffic light is green, the actual delay at this intersection is zero; if the encountered traffic light is red, the actual delay at this intersection equals the remaining time of the red phase.

*4.1.3. Passengers*

We assume that there are four types of passenger arrival rates. The passenger generating rate at bus stop $e$ is denoted by $r_e$ (measured by passengers per minute). Table 4 lists the serial numbers of stops associated with the given arrival rates. Passengers will be generated at all bus stops by the Monte-Carlo method based on the corresponding passenger arrival rates.

Table 4

Bus stops with given arrival rate.

| $r_e$ | Related Stops |
|---|---|
| 1 | 2,4,6,8,13,16,17,19,24,26,28 |
| 2 | 1,3,5,7,10,12,14,15,18,21,22,23,25 |
| 3 | 9,11,27,29 |
| 4 | 20,30 |

Note. $r_e$ is measured by passengers per minute.

Table 5

The probabilities for the following downstream stops to be chosen as destination.

| No. | 1 | 2 | 3 | 4 | 5 | 6 | 7 | 8 | 9 | 10 | 11 | 12 | 13 |
|---|---|---|---|---|---|---|---|---|---|---|---|---|---|
| Series 1 | 0.0135 | 0.027 | 0.0541 | 0.0811 | 0.1081 | 0.1351 | 0.1351 | 0.1216 | 0.1216 | 0.0811 | 0.0541 | 0.0405 | 0.0270 |
| Series 2 | 0.0345 | 0.0862 | 0.1207 | 0.1552 | 0.1724 | 0.1552 | 0.1207 | 0.0862 | 0.0517 | 0.0172 | / | / | / |

To choose destination stops for the new generated passengers, we will use the two series in Table 5. Assume that bus stop $e$ is associated with series 1. The $n$ th element of series 1 is the probability of the $n$ th stop downstream of $e$ to be chosen as a destination stop for the newly generated passengers at $e$. The sum of the elements in any one series is 1. These two series can be regarded as a discrete probability distribution. We assume that series 1 is adopted by bus stops including 1, 7, 10, 13, 20, 21, 27, 30 and series 2 by the other remaining stops.

To evaluate the level of bus service from the point of view of passengers, we will calculate indices including the average waiting time, the average riding time and the average travel time denoted by $t_P^W$, $t_P^R$ and $t_P^{Tr}$, respectively. $P$ stands for the passengers who have finished their bus trips during the observation period. $\sigma_P^W$, $\sigma_P^R$ and $\sigma_P^{Tr}$ are the standard deviations of $t_P^W$, $t_P^R$ and $t_P^{Tr}$, respectively. $n_P$ is the number of passengers in $P$.

*4.1.4. Other Basic Coefficients*

The other commonly used coefficients in this section are given as follows. The length of one observation period $T$ is 4 hours which equals 14400 seconds (s). The discount rate $\gamma$ is set to 0.5. The control action set $A$ commonly used in all control points is {0, 2, 4, 6, 8, 10}. The number of stages to be looked ahead in our strategy will be 3 if no otherwise mentioned.

Since we are dealing with a stochastic system, in order to make the results and the subsequent comparisons more reliable, we will run the simulation system 50 times and then use the average values over these rounds of simulations in the following tables. One round of simulation spans 4 hours, i.e. one observation period.

For simplicity, we assume that all the time quantities, such as the stability index, holding time, and average waiting time, are measured by seconds in the subsequent tables unless otherwise specified.

**4.2. Test the Strategy**

### 4.2.1. Preliminary Impression

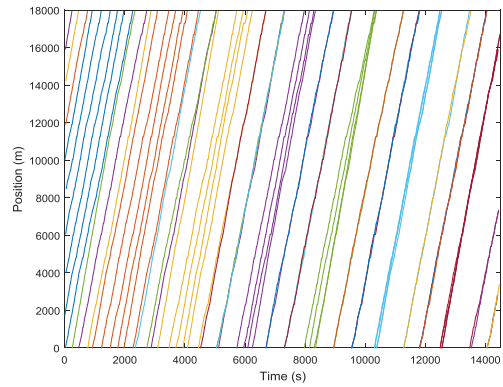

Fig. 2 The trajectories of buses without control.

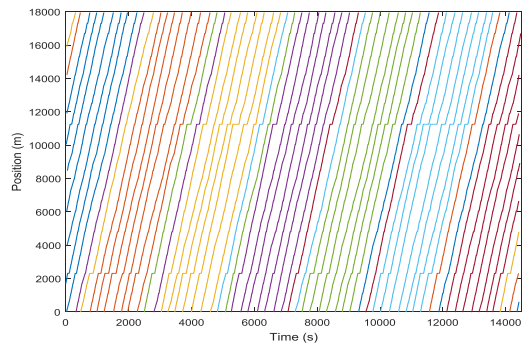

Fig. 3 The trajectories of buses with the terminal station control at stops 5 and 20.

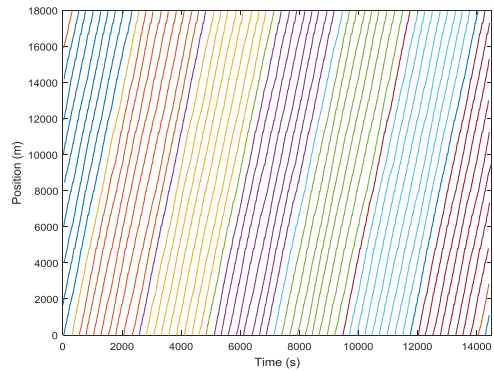

Fig. 4 The trajectories of buses resulted from 3SLA.

To have a preliminary impression about the character of the bus line, Fig. 2 presents the trajectories of buses resulted from the non-control scheme. From these trajectories, we can see that bus bunching happens and becomes more and more serious with time elapsing.

Fig. 2 gives a typical control result under the Terminal Station Holding Strategy (TSHS) which carries out the holding operation at stops 5 and 20. The operational rule of TSHS is as follows. When a bus finishes its boarding and alighting operations at a terminal station, we can compare its forward headway with the ESH and then decide on the holding time. If the forward headway is bigger than the ESH, the bus should be released at once; or else, the bus should be held for a time interval that equals the difference of the ESH minus the forward headway. Under TSHS, bus bunching is removed. The ESH used is 234.65 seconds.

Fig. 3 shows a typical result under our new strategy with 3 stages to be looked ahead. To simplify subsequent expressions, we will use 'nSLA' to denote our new holding strategy with $n$ stages to be looked ahead. The trajectories in Fig. 3 show a very smooth and stable running state of the bus line.

*4.2.2. Influence of Number of Stages to Be Looked Ahead*

Table 6

The stability indices and control actions with respect to different strategies.

| Methods | $\overline{c}_H$ | $\sigma_{\overline{c}}$ | $n_{\overline{T}}$ | $a_\Sigma$ | $\overline{a}$ | $\sigma_{\overline{a}}$ | Bunch |
|---|---|---|---|---|---|---|---|
| No Control | 349.0 | 184.74 | 1768 | / | / | / | Yes |
| TSHS | 47.27 | 15.12 | 1712 | 5397 | 3.15 | 67.54 | No |
| 1SLA | 21.19 | 6.11 | 1695 | 6926 | 4.09 | 4.47 | No |
| 2SLA | 19.46 | 5.49 | 1694 | 6872 | 4.05 | 4.41 | No |
| 3SLA | 17.88 | 5.31 | 1695 | 6764 | 3.99 | 4.40 | No |
| 4SLA | 20.10 | 6.06 | 1696 | 7135 | 4.21 | 4.44 | No |
| 5SLA | 20.74 | 6.88 | 1693 | 7298 | 4.31 | 4.52 | No |

Table 7

Waiting and riding times of passengers with respect to different strategies.

| Methods | $n_P$ | $t_P^W$ | $\sigma_P^W$ | $t_P^R$ | $\sigma_P^R$ | $t_P^{Tr}$ | $\sigma_P^{Tr}$ |
|---|---|---|---|---|---|---|---|
| No Control | 12922 | 327.1 | 240.4 | 426.7 | 191.2 | 753.8 | 316.6 |
| TSHS | 13282 | 131.8 | 80.57 | 433.5 | 197.4 | 565.3 | 211.0 |
| 1SLA | 13167 | 125.2 | 73.4 | 441.0 | 197.2 | 566.2 | 210.1 |
| 2SLA | 13146 | 125.4 | 73.6 | 439.1 | 197.3 | 564.5 | 210.0 |
| 3SLA | 13218 | 123.8 | 73.0 | 435.2 | 197.3 | 559.0 | 209.2 |
| 4SLA | 12918 | 124.5 | 73.2 | 440.7 | 197.9 | 565.2 | 212.3 |
| 5SLA | 13030 | 125.5 | 74.5 | 445.1 | 199.9 | 570.6 | 214.4 |

In Tables 6 and 7, we present the computational results with respect to different strategies. With the holding interference, all the holding strategies can resist bus bunching effectively in the 4 hours observation period. According to the data in Table 6, the TSHS has a relatively big stability index and its holding interference changes more abruptly. Comparing with the series of our 'nSLA' strategies, TSHS has a similar average travel time per passenger. But the average waiting time per passenger resulted from TSHS is higher than from our new strategy. The average riding time per passenger resulted from TSHS is relatively low comparing with our new strategy.

From the data in Table 6, we can observe a trend of the influence of the number of stages to be looked ahead in our new strategy. As the number of stages to be looked ahead increases, the performance resulted from our new strategy improves at the beginning and then changes to a undesirable direction with respect to the stability index and the strength of the outside interference. The data in Table 7 show a similar but relatively weak trend as above. This observation tells us that the number of stages to be looked ahead in our strategy should be carefully selected. A too small or too big number of stages to be looked ahead will reduce the effectiveness of our new strategy.

*4.2.3. The Influence of Different Action Sets*

Table 8

Control action sets with different elements.

| Action Set | Holding times (seconds) |
|---|---|
| 1 | 0,2,4,6,8,10 |
| 2 | 0,3,6,9,12,15 |
| 3 | 0,2,4,6 |
| 4 | 0,5,10,15 |
| 5 | 0,1,2,3,4,5,6,7,8,9,10 |
| 6 | 0,1,2,3,4,5,6,7,8,9,10,11,12,13,14,15 |

Now let us investigate the influence of the feasible holding times on the performance resulted from our strategy. In Table 8, we list 6 different action sets. These sets have different ranges and intervals between two neighboring actions. For example, the first set {0, 2, 4, 6, 8, 10} has a range 10 seconds and an interval 2 seconds. The fourth set {0, 5, 10, 15} has a range 15 seconds and an interval 5 seconds. The computational results with respect to these action sets are summarized in Tables 9 and 10.

Table 9

The stability indices and control actions regarding different action sets.

| Action Set | $\overline{c}_H$ | $\sigma_{\overline{c}}$ | $n_{\overline{T}}$ | $a_\Sigma$ | $\overline{a}$ | $\sigma_{\overline{a}}$ | Bunch |
|---|---|---|---|---|---|---|---|
| 1 | 17.88 | 5.31 | 1695 | 6764 | 3.99 | 4.40 | No |
| 2 | 15.75 | 4.88 | 1680 | 8060 | 4.80 | 5.74 | No |
| 3 | 37.80 | 16.84 | 1723 | 4963 | 2.88 | 2.88 | Yes/No |
| 4 | 16.29 | 5.45 | 1673 | 8226 | 4.92 | 5.92 | No |
| 5 | 21.22 | 6.38 | 1700 | 6572 | 3.87 | 4.11 | No |
| 6 | 16.66 | 5.27 | 1674 | 8030 | 4.80 | 5.70 | No |

Table 10

Waiting and riding times of passengers regarding different action sets.

| Action Set | $n_P$ | $t_P^W$ | $\sigma_P^W$ | $t_P^R$ | $\sigma_P^R$ | $t_P^{Tr}$ | $\sigma_P^{Tr}$ |
|---|---|---|---|---|---|---|---|
| 1 | 13218 | 123.8 | 73.0 | 435.2 | 197.3 | 559.0 | 209.2 |
| 2 | 13069 | 127.3 | 75.6 | 440.8 | 200.8 | 568.1 | 214.2 |
| 3 | 13167 | 127.4 | 75.5 | 433.0 | 194.4 | 560.4 | 208.9 |
| 4 | 13199 | 125.8 | 74.2 | 444.7 | 200.8 | 570.5 | 215.6 |
| 5 | 13134 | 126.4 | 73.5 | 437.9 | 198.2 | 564.3 | 213.3 |
| 6 | 13384 | 123.3 | 73.9 | 444.0 | 197.9 | 567.3 | 211.8 |

According to their structural features, we can conduct the following comparisons: comparing the data associated with sets 1 and 2; comparing the data associated with sets 3 and 4; and comparing the data associated with sets 5 and 6. Based on the comparisons, we can obtain some interesting observations. From the data in Table 9, we can see that in the pairs of sets mentioned above, the latter outperforms the former with respect to the stability index. To realize the above superiority, the latter requires stronger outside interference, i.e. bigger average holding time, than the former. For example, the stability index associated with set 2 is 15.75 seconds that accompanies with the average holding time 4.8 seconds. But the stability index associated with set 1 is 17.88 seconds that accompanies with the average holding time 3.99 seconds.

From the data in Table 10, we can observe another phenomenon associated with these set pairs. The former in the pair outperforms the latter with respect to the average riding time per passenger and the average total travel time per passenger. For example, the average riding time and the average total travel time associated with set 1 are 435.2 seconds and 559.0 seconds, respectively. The average riding time and the average total travel time associated with set 2 are 440.8 seconds and 568.1 seconds, respectively.

The above two observations can be explained as follows. Two compared sets have the same number of holding times. The one with the bigger range can regulate the operation of the bus line with relatively strong interference. But to obtain a better performance, the cost indicated by the average holding time is relatively high for the action set with the bigger range. The average riding time is directly influenced by the size of the average holding time. If the influence of different action sets on the average waiting time is relatively small, the average total travel time should have a similar changing trend as the average riding time.

If we observe the data resulted from sets 2, 4 and 6, we can obtain another interesting observation. Though the ranges of these three sets are the same, their intervals are different. The intervals of sets 2, 4 and 6 are 3 seconds, 5 seconds and 1 second, respectively. The performances associated with these three sets show that a too small or too big interval of an action set generally reduces the performance associated with the action set regarding the stability index.

Another important observation from the data in Tables 9 and 10 is that if the range and the interval of an action set, e.g. the third set {0, 2. 4. 6}, are too small, the corresponding strategy may fail to resist bus bunching. So when a high-frequency bus line with strong trend to bunching is under consideration, the range and interval of the actually used action set should not be too small.

*4.2.4. Influence of the Number of Control Points*

Now let us consider the influence of the number of control points on the performance of the bus line under our new strategy. In Table 11, we list 5 sets of control points.

Table 11

The sets of bus stops which are used as control points.

| Sets of Stops | The corresponding Bus stops |
|---|---|
| 11BS | 2,3,5,11,15,16,17,20,21,25,29 |
| 9BS | 2,5,11,15,16,20,21,25,29 |
| 7BS | 2, 11,15,16,20,25,29 |
| 5BS | 11,15,16,20, 25 |
| 3BS | 11,16, 25 |

In Tables 12 and 13, we present the computational results with respect to the sets in Table11. In Table 12, we can observe an obvious trend that the stability index, the average holding time and the total outside interference will gradually increase with the decrease of the number of control points. This observation is corresponding to our intuition that the more control points are used, the more powerful the holding strategy will be.

From the data in Table 13, we can see that various average times per passenger do not show a changing trend similar to the one observed from the data in Table 12. The differences among the average waiting times or the average riding times resulted from different sets of control points are relatively small.

Table 12

Stability indices and control actions regarding different sets of control points.

| Sets of Stops | $\bar{c}_H$ | $\sigma_{\bar{c}}$ | $n_{\bar{T}}$ | $a_\Sigma$ | $\bar{a}$ | $\sigma_{\bar{a}}$ | Bunch |
|---|---|---|---|---|---|---|---|
| 11BS | 17.88 | 5.31 | 1695 | 6764 | 3.99 | 4.40 | No |
| 9BS | 18.39 | 5.08 | 1696 | 6991 | 4.12 | 4.45 | No |
| 7BS | 19.69 | 6.15 | 1697 | 7004 | 4.13 | 4.42 | No |
| 5BS | 21.28 | 7.33 | 1691 | 7020 | 4.15 | 4.41 | No |
| 3BS | 23.67 | 6.46 | 1688 | 7140 | 4.23 | 4.50 | No |

Table 13

Waiting and riding times of passengers with respect to different sets of control points.

| Set of Stops | $n_{P_1}$ | $t_{P_1}^W$ | $\sigma_{P_1}^W$ | $t_{P_1}^R$ | $\sigma_{P_1}^R$ | $t_{P_1}^{Tr}$ | $\sigma_{P_1}^{Tr}$ |
|---|---|---|---|---|---|---|---|
| 11BLS | 13218 | 123.8 | 73.0 | 435.2 | 197.3 | 559.0 | 209.2 |
| 9BLS | 13218 | 125.5 | 73.8 | 440.8 | 196.4 | 566.4 | 210.0 |
| 7BLS | 13162 | 125.2 | 73.7 | 440.8 | 195.9 | 566.1 | 210.1 |
| 5BLS | 13068 | 124.8 | 73.1 | 437.0 | 196.5 | 561.8 | 211.0 |
| 3BLS | 13284 | 125.6 | 73.6 | 440.8 | 197.5 | 566.4 | 211.2 |

**4.3. Computational Time**

Table 14

The computational times associated with the implementation of our new strategy.

| Number of stages to be looked ahead | 1 | 2 | 3 | 4 | 5 |
|---|---|---|---|---|---|
| Time to run the simulation once (sec) | 0.203 | 0.828 | 4.437 | 24.268 | 144.828 |
| Time to make one decision (sec) | <0.001 | <0.001 | 0.015 | 0.016 | 0.094 |

Table 14 presents the related computational times associated with the implementation of our new strategy. The observation time period is still four hours. The control points constitute the first set in Table 11. The action set associated with any control point is the first set in Table 8.

The program is realized with JAVA language in the NetBeans IDE 8.0.2. The computer used has a processor of Intel® Core i3-3120M CPU @2.50GHz. The installed memory (RAM) of the computer is 4.00GB (2.32GB usable).

Because in NetBeans IDE, there is no effective language tool which can used to distinguish a time interval smaller than a millisecond, we will use '<0.001' to replace the computational time which is less than a millisecond in Table 14. The resulted computational times are very promising for the application of the new approach in practice.

## 5. Conclusion

To avoid choosing a myopic holding time, we proposed a new headway-based holding strategy based on the multi-stage looking-ahead mechanism.

To supply a criterion for adjusting bus headway especially when we deal with a high-frequency bus line without any pre-specified schedule or headway, we define the DTH associated with a CTP as the average time headway at the CTP. Based on DTH, we further define the stability index of a bus line and the control cost associated with a specified holding time. We can evaluate the whole stability of a bus line with the stability index and choose a proper holding time by directly minimizing the corresponding control cost. Our strategy can make use of the global headway information through the above operations.

We introduced the multi-stage looking-ahead mechanism into the selection of a proper holding time. Using the multi-stage looking-ahead mechanism, we can assess the influence of a chosen holding time on the operational stability over a relatively long time period. By doing so, we can further avoid making a possibly myopic decision. Since we realized the multi-stage looking-ahead mechanism in our strategy with a detailed simulation system, we can easily deal with all kinds of complex factors influencing a bus line, such as the signalized intersections and the random arrivals of passengers at stops.

The numerical experiment demonstrated the effectiveness of our new strategy. Some meaningful insights were uncovered by the numerical analyses. They are summarized as follows:

- To obtain the optimal performance, we should select the number of stages to be looked ahead in our strategy carefully. The number of stages should not be too small or too big because they usually give rise to a relatively low performance. By the way, to look ahead more stages will be time consuming.
- The bigger the range of an action set is, the better the performance associated with the action set will be.
- The interval between two neighboring feasible holding times in an action set should be determined carefully. If the interval is too small, it is time consuming to find out the optimal holding time from an action set with many elements. If the interval is too big, we may miss out the optimal holding time.
- Generally speaking, the more stops are used as control points, the better the performance of our new holding strategy will be.

Several related research directions can be further pursued in the future. a). In real life, due to various reasons, it is not every stop fit for carrying out the holding operation. In this situation, we need to answer the following questions: which stops should be chosen and how many stops should be used as control points? b). It is possible to extend the multi-stage looking-ahead mechanism to other types of strategies, e.g. the strategies based on traffic signal adjustment. c). In the future, we may consider how to stabilize multiple bus lines at the same time. d). Holding buses at control points will change the original pollution emission from vehicles. We need to find out what is the specific influence of serial holding operations on the traffic pollution emission.

**Acknowledgment**

This research was supported in part by National Natural Science Foundation of China(71601118, 71801153, 71871144），the Natural Science Foundation of Shanghai(18ZR1426200).